\documentclass[runningheads]{llncs}
\usepackage[T1]{fontenc}
\usepackage{graphicx}
\usepackage{float}
\usepackage{hyperref}
\usepackage{amsmath,amssymb}
\usepackage{algorithm}
\usepackage[noEnd=true, indLines=true]{algpseudocodex}
\makeatletter
\algdef{SE}[KERNEL]{Kernel}{EndKernel}{%
  \algpx@startCodeCommand\algpx@startIndent\textbf{launch} GPU kernel%
}{}
\algtext*{EndKernel}
\pretocmd{\Kernel}{\algpx@endCodeCommand}{}{}
\pretocmd{\EndKernel}{\algpx@endIndent}{}{}
\pretocmd{\EndKernel}{\algpx@endCodeCommand[1]}{}{}
\algdef{SE}[KERNELB]{KernelBatched}{EndKernelBatched}{%
  \algpx@startCodeCommand\algpx@startIndent\textbf{launch} batched GPU kernel(s)%
}{}
\algtext*{EndKernelBatched}
\pretocmd{\KernelBatched}{\algpx@endCodeCommand}{}{}
\pretocmd{\EndKernelBatched}{\algpx@endIndent}{}{}
\pretocmd{\EndKernelBatched}{\algpx@endCodeCommand[1]}{}{}
\makeatother

\usepackage{subcaption}

\usepackage{changes} 
\definechangesauthor[name={FC}, color=orange]{FC}
\definechangesauthor[name={RP}, color=green]{RP}
\definechangesauthor[name={TC}, color=purple]{TC}

\begin{document}
\title{Accelerating the Solving of Many Tiny General Linear Systems on GPUs: Application to Constitutive Laws}
\titlerunning{Solving of Many Tiny General Linear Systems on GPUs}
%
\author{
Tristan Chenaille\inst{1,2} \and
Francesca Cuteri\inst{1} \and
Raphaël Prat\inst{1} \and
Guillaume Latu\inst{1} \and
Thomas Helfer\inst{1}
}
\authorrunning{T. Chenaille et al.}
%
\institute{CEA, DES, IRESNE, DEC, SESC, Cadarache, St-Paul-lez-Durance, France \and
Aix-Marseille University, Mathematics and Computer Science Doctoral School, France \\
\email{tristan.chenaille@cea.fr}}
\maketitle              
\begin{abstract}
Many applications require solving large numbers of independent linear systems on GPUs. While this need is well addressed for small to large systems, tiny ones, understood here as systems of dimension below 32, remain challenging. This is especially relevant in constitutive law evaluation, where millions of integration points are handled independently, and where each constitutive update generally relies on a Newton iterative method. Each iteration then requires the double-precision solution of a tiny general square linear system using LU factorization with partial pivoting (LUpp). The present study was conducted within a closed-source prototype, which serves as a demonstrator for porting to NVIDIA GPUs constitutive law evaluations currently provided on CPUs by TFEL/MFront, an open-source code generation tool for material knowledge. We compare several double-precision LUpp solvers, including implementations from GPU linear algebra libraries as well as custom-designed CUDA kernels. The comparison is performed first on large batches of standalone linear systems, and then within the full constitutive-law evaluation workflow, where each integration point requires a sequence of distinct linear systems, one per iteration of its own Newton loop. The study shows that the best LUpp solving strategy strongly depends on several factors including system size and application context. We discuss several key aspects, including register pressure, occupancy, the ability to invoke the LUpp solver directly from device code, and whether assigning several threads to each system is the most efficient strategy. Experiments on an NVIDIA H100 GPU show that the specialized LUpp solvers proposed in this work can outperform existing state-of-the-art approaches for this class of workloads, with speedups of up to $6.5\times$ over cuSolverDx, and up to $17.7\times$ over MAGMA.

\end{abstract}

\newpage

\section{Introduction}
\label{sec:introduction}

Graphics Processing Units (GPUs) have become a key component of modern high-performance computing systems due to their ability to accelerate many workloads. This trend has shaped the exascale era, with the main exascale systems relying heavily on GPU-based acceleration~\cite{ref_gpu_exascale_era}. As a result, porting traditional CPU-oriented scientific applications to GPUs is now essential. Achieving high performance, however, is not straightforward and often requires substantial algorithmic and data structures rework. Attainable performance also depends on whether the underlying problem can expose enough fine-grained parallelism. Applications involving large numbers of independent computations, including the solution of many linear systems, are typically well suited to GPU execution.

GPU-based linear algebra routines have been extensively developed, and efficient implementations are now available for a wide range of problem sizes. In particular, many solvers have demonstrated high performance for small to large linear systems. However, a different regime arises when considering tiny systems, i.e. those of dimension $n \leq 32$. The amount of computation within these systems is limited, which reduces the available parallelism and makes it difficult to fully utilize GPU capabilities. Thus, overheads and implementation details that are negligible for larger systems can become dominant, and strategies that perform well at larger scales do not necessarily remain efficient~\cite{ref_magma_larger_strategies_do_not_remain_efficient}. This specialized regime calls for dedicated optimization strategies, especially in applications where such systems must be solved in large numbers. 

Such workloads arise in several fields, including reactive flow simulations~\cite{ref_sundials_many_odes}, block-Jacobi preconditioning~\cite{ref_block_jacobi_lu}, and computational solid mechanics. In this work, we focus on the latter, and more specifically on evaluating stress-strain constitutive models at integration points. This task is currently performed on CPUs using the TFEL/MFront library~\cite{ref_tfel_mfront}, which compiles a high-level description of the constitutive law. To load the latter, global simulation codes use the MFrontGenericInterfaceSupport (MGIS) library~\cite{ref_mgis}, which drives the behavior evaluation. A representative use case is the high-performance simulation of nuclear fuel elements within the PLEIADES framework~\cite{ref_pleiades}: in the MFEM-MGIS application~\cite{ref_mfem_mgis},
%
to model the response of fuel under irradiation, phenomenas such as viscoplasticity, swelling, and damage are captured by constitutive models.

In large-scale simulations, the material response must be computed at a vast number of integration points, each processed independently. At each point, it is typically obtained by solving a nonlinear problem using a Newton iterative method. Each iteration requires the solution of a general linear system using LUpp in double precision. Although each individual system is small to tiny (usually $n\leq20$), the very large number of independent integration points can make the overall cost significant, which makes this workload a natural candidate for GPU acceleration. It can also avoid costly host-device data transfers: when the surrounding global simulation runs on the GPU, keeping the constitutive evaluation on the CPU would require moving integration-point data between host and device at every iteration. Porting constitutive model evaluation to GPUs is therefore a step toward faster simulations, better hardware utilization, and a reduced energy footprint.

In this work, we investigate the efficient solution of tiny general linear systems using LUpp on GPUs, with a particular focus on application-driven workloads arising from constitutive law evaluation. We study behavior integration as a batch of integration points on the GPU, decoupled from the surrounding global solution procedure. This focuses the comparison on the LUpp solving strategy and its interaction with the constitutive update, rather than on orchestration choices made by the global solver. When behavior integration is instead fused into a global assembly loop, the conclusions may differ. Our contributions are threefold. First, we design specialized LUpp solvers tailored to tiny systems, exploring different implementation strategies. These solvers are device-callable, meaning that they can be invoked directly from device code. Second, we provide a comprehensive comparison of these solvers against existing implementations from external libraries, both on large batches of standalone linear systems and within the full constitutive law evaluation workflow. Third, experimental results obtained on an NVIDIA H100 GPU demonstrate that the optimal approach strongly depends on both the system size and the application context, and that the proposed specialized LUpp solvers can outperform existing state-of-the-art ones by up to $17.7\times$ in specific cases.

\section{Background and Related Work}
\label{sec:background_and_related_work}

In this section, we briefly recall principles of Newton--Raphson method and LU-based solving, both of which are central to constitutive law evaluation under implicit schemes. We then review existing approaches for solving tiny linear systems on GPUs, and their adoption in constitutive modeling libraries.

\subsection{Newton-Raphson Iterative Method}

The Newton--Raphson method is a standard approach for solving nonlinear equations. Under standard assumptions, the method exhibits quadratic convergence when the iterate is sufficiently close to the solution~\cite{ref_newton_convergence}. \\ 
In the vectorial case, given a nonlinear system
\begin{equation}
\mathbf{F}(\mathbf{x}) = \mathbf{0}, \quad \text{with} \:\: \mathbf{x} \in \mathbb{R}^n \:\: \text{ and } \:\: \mathbf{F} : \mathbb{R}^n \to \mathbb{R}^n,
\end{equation}
At iteration $k$, the method solves (typically using LUpp) the linearized system
\begin{equation}
\mathbf{J}(\mathbf{x}^{(k)}) \, \Delta \mathbf{x}^{(k)} = -\mathbf{F}(\mathbf{x}^{(k)}),
\end{equation}
where $\mathbf{J}(\mathbf{x}) = \partial \mathbf{F} / \partial \mathbf{x}$ is the square, general Jacobian matrix of $\mathbf{F}$, $\Delta \mathbf{x}^{(k)}$ is the correction, and $\mathbf{F}(\mathbf{x}^{(k)})$ is the residual. The solution is then updated as
\begin{equation}
\mathbf{x}^{(k+1)} = \mathbf{x}^{(k)} + \Delta \mathbf{x}^{(k)}.
\end{equation}
Iterations stop when a prescribed convergence criterion is met, typically based on the residual norm, the correction norm, or a maximum number of iterations.

\subsection{LU Factorization, Triangular Solves, and Variants}

LU factorization, also referred to as LU decomposition, consists in factorizing a square, general matrix $A$ as the product of a lower triangular matrix $L$ and an upper triangular matrix $U$. This is achieved through Gaussian elimination, which proceeds column by column: at each step, the current pivot row is used to eliminate entries below it, producing a reduced submatrix known as the trailing submatrix or Schur complement, on which the process is applied recursively. LU factorization is widely used to solve linear systems of the form $Ax=b$, since, once the $LU$ factorization of $A$ is available, the solution can be obtained by solving two simple triangular systems successively, namely $Ly=b$ and $Ux=y$. Numerical stability and robustness can be respectively improved and guaranteed by performing row interchanges during Gaussian elimination so that the pivot element is nonzero and as large as possible in magnitude. This strategy is known as partial pivoting. With partial pivoting, the factorization is written $PA=LU$, where $P$ is a permutation matrix encoding row interchanges. This pivoting technique is widely used; while guaranteeing robustness, it offers a good compromise between numerical stability and computational cost~\cite{ref_gepp_standard}.

The factorization step dominates the computational cost, requiring $\mathcal{O}(n^3)$ floating-point operations (flops). Partial pivoting introduces an additional $\mathcal{O}(n^2)$ comparison cost. The two triangular solves require $\mathcal{O}(n^2)$ flops.

From an algorithmic point of view, LU factorization is sequential across elimination steps, since each step depends on the previous ones. Parallelism can therefore only be exploited within the operations associated with each step. Different variants have been proposed to organize these operations~\cite{ref_various_lu_golub}, and can be grouped in two families. In right-looking (RL) variants, the transformations computed at a given elimination step are immediately applied to the trailing submatrix. This exposes substantial matrix-matrix parallelism and maps well to BLAS-3 operations. In left-looking (LL) variants, the current step is instead updated by reusing transformations computed during previous steps. Although they generally expose less immediate matrix-matrix parallelism than RL variants, they can reduce repeated accesses to matrix entries. Orthogonally to this scheduling choice, blocked variants partition the matrix to exploit data locality. The matrix partitioning is often one-dimensional, grouping consecutive columns into panels, but can also be two-dimensional, dividing the matrix into a grid of tiles~\cite{ref_tiled_lu_buttari}. All these design choices have a strong impact on data locality, available parallelism, and ultimately performance on different architectures.

\subsection{State-of-the-Art GPU Optimization Strategies for Tiny LUpp} \label{subsec:sota_tiny}
The scientific community has studied the efficient GPU implementation of batched partial-pivoting LU (LUpp), from which key optimization principles for the tiny regime ($n \leq 32$) can be identified. These principles are largely shared with QR and Cholesky~\cite{ref_magma_larger_strategies_do_not_remain_efficient}. A central observation is that the factorization is largely memory-bound: data transfer costs dominate the $\mathcal{O}(n^3)$ arithmetic~\cite{ref_guide,ref_magma_larger_strategies_do_not_remain_efficient}. The classical strategy for large matrices relies on blocked algorithms that partition the matrix into panels and cast the trailing-matrix update as Level-3 BLAS to approach peak compute throughput. For tiny sizes, this approach becomes ineffective: the updates are too small to amortize their overhead~\cite{ref_progressive_optimization}. Optimizing performance in this regime therefore requires size-aware kernel designs that minimize data access and movement.

The thread-to-system mapping, i.e.\ how GPU threads are organized to process each independent system, is a central design choice. In the one-thread-per-system approach, a single thread processes an entire matrix typically held in its private register file, which eliminates all synchronization. This paradigm is generally used only for the smallest sizes of the tiny regime ($n\leq8$): each thread must store all $n^2$ elements plus many temporaries, so local memory usage grows rapidly with $n$ and degrades performance~\cite{ref_villa_power_performance,ref_high_perf_batched_lu}.

In the multi-thread approach, a group of threads cooperates on one matrix, each thread owning one row~\cite{ref_magma_larger_strategies_do_not_remain_efficient,ref_guide}. By spreading the matrix across threads, this design lowers per-thread register pressure, which usually avoids using local memory and increases multiprocessor occupancy~\cite{ref_predictive_model}. Two storage strategies then arise. In the first one, the entire matrix is stored in shared memory, so every arithmetic operation depends on a shared-memory access. In the second one, register-blocking is used: each thread permanently holds its row in registers~\cite{ref_guide}, with inter-thread communication relying on warp shuffles or on a small shared memory workspace. This significantly reduces data traffic. The LL and RL variants are then preferable in different LUpp cases: with the matrix in shared memory, the LL one reduces the number of accesses and the on-chip data movement~\cite{ref_guide}, whereas with register-blocking, the RL one exposes more per-step parallelism and simplifies the LU's partial pivoting.

Several other optimizations are mentioned in the literature. Dedicated data layouts preserve coalesced memory accesses~\cite{ref_guide,ref_batched_matrix_computations}. Tunable concurrency assigns multiple factorizations to a single thread group, increasing per-block workload~\cite{ref_magma_larger_strategies_do_not_remain_efficient,ref_factorization_million_matrices}. Kernel fusion combines the factorization, triangular solves, and potentially application-level operations into a single kernel, eliminating redundant global memory round-trips~\cite{ref_progressive_optimization,ref_factorization_million_matrices}. Compile-time specialization via C++ templates enables loop unrolling, optimized register allocation, and aggressive instruction scheduling by the compiler~\cite{ref_predictive_model,ref_magma_larger_strategies_do_not_remain_efficient}. Finally, specific to LUpp, logical pivoting records pivot indices in a vector and avoids the physical row interchange at each elimination step~\cite{ref_magma_larger_strategies_do_not_remain_efficient,ref_factorization_million_matrices}. Our prototype builds on all these principles.

\vspace{-0.3cm}
\subsection{Existing GPU Libraries for Tiny LUpp Solvers}
\label{existing_gpu_libraries_for_tiny_lupp}

Leveraging some of these insights, a number of libraries implement LUpp on NVIDIA GPUs for solving large numbers of independent linear systems. These implementations can be divided into two categories. On the one hand, batched solvers, such as those provided in MAGMA~\cite{ref_magma} and cuBLAS~\cite{ref_cublas}, operate on large sets of systems but are typically invoked from the CPU. This forces the linear solve to be executed as a separate GPU phase, rather than being fused with application computations, which introduces additional overheads~\cite{ref_kernel_fusion}. On the other hand, libraries such as cuSolverDx~\cite{ref_cusolverdx}, KokkosKernels~\cite{ref_kokkoskernels}, and SNLS~\cite{ref_snls} provide device-callable solvers that can be invoked directly from within GPU kernels. This approach avoids host-device synchronization and enables tighter integration within application-driven GPU workflows.

Within GPU-based constitutive law evaluation, whether provided by external libraries or developed in-house, linear solvers must be considered within a broader pipeline involving application-level constraints such as Newton iterations and constitutive updates. Figure~\ref{fig:lupp_paradigms_algo} shows how the batched and device-callable approaches shape this workflow: the former leads to a CPU-driven evaluation, the latter to a fully GPU-driven one. The CPU-driven scheme breaks each Newton iteration into several kernel launches. This adds launch overheads, moves intermediate states through global GPU memory, and requires additional synchronization for convergence checks, which transfer only a single flag. The GPU-driven scheme avoids these costs by keeping the Newton loop inside a single GPU kernel. The next subsection reviews how these two paradigms are adopted in existing constitutive modeling libraries.

\vspace{-25pt}
\begin{figure}[h]
\centering
\begin{minipage}[t]{0.48\linewidth}
\begin{algorithm}[H]
\caption{Batched LUpp, \\ CPU-driven constitutive evaluation}
\label{alg:batched}
\begin{algorithmic}
\State \textbf{send} inputs: CPU $\rightarrow$ GPU
\While{not all points converged}
  \Kernel
    \ForAll{points $i$ \textbf{in parallel}}
      \State assemble $F_i, J_i$ \Comment{Law}
    \EndFor
  \EndKernel
  \KernelBatched
    \ForAll{points $i$ \textbf{in parallel}}
      \State solve $J_i \Delta x_i = -F_i$ \Comment{LUpp}
    \EndFor
  \EndKernelBatched
  \Kernel
    \ForAll{points $i$ \textbf{in parallel}}
      \State $x_i \gets x_i + \Delta x_i$
      \State check Newton convergence
    \EndFor
  \EndKernel
  \State \textbf{send} convergence: GPU $\rightarrow$ CPU
\EndWhile
\State \textbf{send} outputs: GPU $\rightarrow$ CPU
\end{algorithmic}
\end{algorithm}
\end{minipage}
\hfill
\begin{minipage}[t]{0.48\linewidth}
\begin{algorithm}[H]
\caption{Device-callable LUpp, GPU-driven constitutive evaluation}
\label{alg:devicecallable}
\begin{algorithmic}
\State \textbf{send} inputs: CPU $\rightarrow$ GPU
\Kernel
  \ForAll{points $i$ \textbf{in parallel}}
    \While{point not converged}
      \State assemble $F_i, J_i$ \Comment{Law}
      \State solve $J_i \Delta x_i = -F_i$ \Comment{LUpp}
      \State $x_i \gets x_i + \Delta x_i$
      \State check Newton convergence
    \EndWhile
  \EndFor
\EndKernel
\State \textbf{send} outputs: GPU $\rightarrow$ CPU
\end{algorithmic}
\end{algorithm}
\end{minipage}
\caption{Simplified comparison of GPU-based constitutive law evaluation with a batched versus a device-callable LUpp solver (no tangent operator computation).}
\label{fig:lupp_paradigms_algo}
\end{figure}

\vspace{-30pt}
\subsection{GPU-based Implicit Constitutive Law Evaluation}

In computational mechanics, and in particular in constitutive law integration, implicit schemes based on Newton iterative methods are widely used for their numerical efficiency and their consistency with the global simulation procedure. Locally, they provide a robust framework for solving nonlinear constitutive equations, while their consistent linearization supplies the tangent operator needed by the global equilibrium solver~\cite{ref_simo_consistent_1985}. In contrast, explicit schemes avoid such local iterative solves but may be subject to stronger stability constraints and are not always suitable in the same settings~\cite{ref_de_souza_neto_plasticity}. As a consequence, implicit constitutive updates typically require the repeated LUpp solution of small to tiny linear systems involving the Jacobian of the local nonlinear constitutive equations, until a prescribed convergence criterion is reached. Since convergence behavior depends on the local state, different integration points may require different numbers of Newton iterations, which can lead to irregular computational patterns in a parallel workflow. In some cases, this can result in significant load imbalance, making efficient GPU execution delicate~\cite{ref_gpu_constitutive_irregular_spectral}.

On the batched side, several libraries and frameworks rely on a formulation in which each Newton step is applied uniformly to all integration points. This is notably the case for NEML2~\cite{ref_neml2}, JAX-CPFEM~\cite{ref_jaxcpfem}, and the FEniCSx external-operator framework together with its associated libraries~\cite{ref_latyshev}. MOOSE can also access this type of GPU path through its NEML2 integration~\cite{ref_moose_neml2}. These works cover a broad range of implicit constitutive models, including viscoplasticity, poroplasticity, and crystal plasticity. A structural cost of this batched paradigm is that integration points that have already converged still participate in subsequent Newton iterations until the last point in the batch reaches convergence. This cost is hard to quantify, as it highly depends on the input data and the constitutive law. Still, the number of Newton iterations across integration points generally follows a light-tailed distribution, which leads the libraries above to consider this cost to be acceptable to pay.

On the device-callable side, other implementations keep the Newton loop local to each integration point and invoke the linear solver directly from within the GPU kernel. This strategy is adopted, for example, in ExaCMech~\cite{ref_exacmech}, which relies on SNLS~\cite{ref_snls}, and in AutoMat~\cite{ref_automat}. It has also been explored within Fierro~\cite{ref_fierro}. MOOSE has also recently initiated a native Kokkos-based effort along similar lines~\cite{ref_moose_kokkos}. Such applications span several classes of implicit constitutive laws, notably viscoplastic and crystal plasticity models. A drawback of this strategy is that it may introduce warp-level divergence. In the batched paradigm, the slowdown is driven by the slowest integration point in the entire batch. Here, the effect is confined to the warp level. It is therefore less penalizing.

Taken together, and to the best of our knowledge, these works indicate that there is no consensus on a single dominant strategy for GPU-based implicit constitutive law evaluation. We therefore consider both batched and device-callable LUpp solvers, described in Section \ref{sec:GPU_LUpp_solvers} and compared in Section \ref{sec:performance_evaluation}.

\vspace{-8pt}
\section{GPU LUpp Solvers and Integration Strategies}
\label{sec:GPU_LUpp_solvers}

\vspace{-0.2cm}
In this section, we describe the LUpp solver families implemented in our prototype. Some come from external libraries, while others are developed in-house. We classify them into three categories: batched solvers, multi-threaded device-callable solvers, and single-threaded device-callable solvers.

\vspace{-8pt}
\subsection{Batched Solvers}

\noindent  At present, two libraries provide batched LUpp solvers for NVIDIA GPUs: cuBLAS, NVIDIA's proprietary library, and MAGMA, which is open source. All these routines assume column-major storage and are multi-threaded: each system of dimension $n\leq32$ is processed by $n$ active GPU threads, one per row, enabling coalesced accesses within each column.  

\vspace{0.1cm}

\noindent$\bullet$ \textbf{cuBLAS:} cuBLAS exposes two separate routines: \texttt{cublasDgetrfBatched}, which performs a RL LUpp factorization, and \texttt{cublasDgetrsBatched}, which performs the subsequent triangular solves. The three operands all reside in shared memory: matrix, right-hand side (RHS), and pivot vector. 
\vspace{0.1cm}

\noindent$\bullet$ \textbf{MAGMA:} In this work, MAGMA refers to \texttt{magma\_dgesv\_batched}, a unified routine combining LUpp factorization and triangular solves. Although MAGMA also provides \texttt{magma\_dgetrf\_batched} and \texttt{magma\_dgetrs\_batched}, the unified \texttt{magma\_dgesv\_batched} is more efficient throughout the tiny regime, dispatching for $n \leq 32$ to a specialized RL, logical-pivoting, register-blocking kernel, named \texttt{dgesv\_batched\_small\_kernel}. It uses register-blocking on the matrix and RHS, with the pivot vector in shared memory.

\vspace{-8pt}
\subsection{Device-callable Solvers}

\subsubsection{Multi-threaded Solvers} \label{multithreaded_solvers}

In the multi-threaded approach, a group of threads cooperates on a single system. This follows the conventional GPU rationale: a single thread's register limit is too low to hold a system without using local memory. The work is distributed across several cooperating threads instead. This lowers per-thread register pressure and can raise occupancy. The matrix is spread over these threads, held in their combined registers or in shared memory, and the threads synchronize to advance the factorization together.

\vspace{0.1cm}

\noindent $\bullet$ \textbf{MAGMA-custom (in-house):} \label{subsec:magmacustomdescription} This solver reimplements MAGMA's internal \texttt{dgesv\_batched\_small\_kernel}. The numerical LUpp solve itself is unchanged: column-major storage, RL arithmetic, logical pivoting, and register-blocking are all kept as is. Only the thread-to-data mapping differs, so numerical results and stability are identical. The reimplementation extends the kernel in three ways.

First, it is device-callable. MAGMA already implements this solve on the device, but the routine is internal to the compiled library and is not declared in any public header. Only its host-side launcher is exposed, so the solve cannot be called from a user kernel. Our version is callable from device code, so it can be fused inside a single Newton kernel.

Second, it packs several systems per block. MAGMA assigns one system per block with $n$ threads. When $n < 32$, this leaves $32 - n$ lanes of the warp idle. We launch full warps instead and fit as many systems as possible. For $n = 12$, a warp solves two systems and wastes only $8$ lanes.

Third, the number of rows per thread $K$ is a tunable template parameter ($1$ to $6$). MAGMA fixes $K = 1$, so a system of size $n$ needs $n$ threads. With $K$ rows per thread a system needs $T = \lceil n / K \rceil$ threads, and a warp holds $\lfloor 32 / T \rfloor$ systems. For $n = 12$ and $K = 2$, six threads solve one system, five systems fit in a warp, and only two lanes stay idle.

\label{subsec:phantom_skip}When $n$ is not a multiple of $K$, a phantom mechanism absorbs the remainder. The tail thread owns dead rows that are skipped at compile time. This adds no computational cost, and no register pressure, since every thread allocates the same $K$ slots. Coalescing is preserved: rows are assigned round-robin, so within a system consecutive threads read consecutive memory along each column.

\vspace{0.1cm}

\noindent $\bullet$ \textbf{cuSolverDx-block:} This solver is the RL \texttt{gesv\_partial\_pivot} routine of cuSolverDx, a proprietary device-callable NVIDIA library. It runs in the \texttt{Block()} execution mode, in which the threads of a block cooperate on the systems assigned to that block. All operands reside in shared memory and can benefit from coalesced accesses. The \texttt{SystemsPerWarp} parameter sets how many systems a warp processes. As with cuBLAS, the internal implementation remains opaque.

\vspace{-0.4cm}
\subsubsection{Single-threaded Solvers}
\label{subsubsec:single_threaded}

\noindent In the single-threaded approach, each GPU thread solves an entire system on its own. The natural design keeps the whole matrix in the thread's registers, but, as discussed in Section~\ref{subsec:sota_tiny}, this is only viable for small $n$ ($n\lesssim8$). The following solvers therefore let their operands be placed in global or shared memory, used here not as a cooperation workspace, as in the multi-threaded approach, but as a private extension of each thread's register space. This reduces the unintended use of local memory. However, for large $n$ ($n \gtrsim 30$ on GPUs with a $227$KB per-block shared-memory limit such as the H100), the shared variant can force the block size below $32$ to fit the shared-memory budget, leaving idle warp lanes and lowering effective thread-level occupancy.

\vspace{0.1cm}

\noindent $\bullet$ \textbf{cuSolverDx-thread:} This solver is the RL \texttt{gesv\_partial\_pivot} routine of cuSolverDx running in the \texttt{Thread()} execution mode. All three operands can be placed in registers, shared memory, or global memory. In shared or global memory, only the RHS can benefit from coalesced accesses. 

\vspace{0.1cm}

\noindent $\bullet$ \textbf{KokkosKernels:} This solver processes each system using two open-source KokkosKernels functions: \texttt{SerialGetrf} and \texttt{SerialGetrs}. The factorization is a recursive RL, physical-pivoting LU, like LAPACK's \texttt{getrf2}. Since device code cannot recurse cheaply, recursion is emulated by a fixed-size per-thread software stack. Passing the operands as strided Kokkos views lets each reside in registers, shared, or global memory, enabling coalesced access in the latter two cases.

\vspace{0.1cm}

\noindent $\bullet$ \textbf{SNLS:} This solver uses SNLS's open-source \texttt{SNLS\_LUP\_Solve} routine, a simple RL, logical-pivoting LUpp. While the pivot vector must stay in registers, matrix and RHS can be placed in registers, shared memory, or global memory. In shared or global memory, neither can benefit from coalesced accesses.

\vspace{0.1cm}

\noindent $\bullet$ \textbf{TFEL-baseline:} This solver is used as the baseline in this work because it corresponds to the open-source \texttt{tfel::math::TinyMatrixSolve} routine, already used in production through the TFEL/MFront/MGIS framework. This routine was originally designed for CPU execution, where it is competitive with reference implementations. Its GPU port used here is direct and includes no GPU-specific optimization. It is a LL, logical-pivoting LUpp solver. All operands can be placed in registers, shared memory, or global memory, enabling coalesced access in the latter two cases through \texttt{tfel::math::StridedCoalescedView} objects.

\vspace{0.1cm}

\noindent $\bullet$ \textbf{Tiled (in-house):} \label{tiledsolverdesign} These solvers come in RL and LL variants. They split the matrix into square tiles of tunable size ($\leq6$ to limit register spilling) and run Gaussian elimination on them rather than on scalar entries. The full matrix stays in shared memory (tiled-SHMEM) or global memory (tiled-DRAM), and only the tiles involved in the current step are loaded into registers. At each step, the diagonal tile is factored using scalar partial pivoting, then surrounding tiles are updated with a Schur complement according to the RL or LL schedule. To keep data local, the pivot row is sought and physically swapped inside that tile, and a lower tile is accessed for logical pivoting only when the best in-tile candidate falls below a tunable threshold. Tile-local pivoting admits slightly more pivot growth than textbook partial pivoting, a stability cost is quantified in Section \ref{pure_lupp_performance}. When the system dimension is not a multiple of the tile size, a phantom mechanism pads the trailing tile to full size and uses compile-time skips (cf Section~\ref{subsec:phantom_skip}). The RHS and pivot vector can independently stay in registers or follow the matrix in shared or global memory, where all operands can benefit from coalesced accesses.

\vspace{-0.3cm}
\subsection{Embedding LUpp Solvers in Constitutive Law Evaluation}

Embedding a single-threaded device-callable LUpp solver is straightforward. The MFront-generated per-integration-point code is wrapped in a single CUDA kernel hosting the whole constitutive law evaluation, within which the only changes are placing the LUpp operands in the residency targeted by the new solver (registers, shared, or global memory, instead of TFEL's default register-resident layout) and substituting the new solver's entry point for the \texttt{TinyMatrixSolve} call. The Newton loop and law evaluation are reused as-is.

A multi-threaded device-callable LUpp solver additionally requires partitioning material variables, Jacobian assembly, and convergence checks across the $T$ threads of each cooperating group, as well as adding synchronization barriers.

Embedding a batched LUpp solver requires the MFront-generated code refactor outlined in Figure~\ref{fig:lupp_paradigms_algo}. The originally fused per-point computation is split into up to ten device kernels per global CPU-driven Newton iteration, interleaved with the batched library calls. On top of it, this scheme pays a host-device synchronization per iteration for the global convergence flag check, and routes intermediate GPU-resident data through global memory across each kernel boundary.

\vspace{-0.3cm}
\section{Performance Evaluation and Discussion}
\label{sec:performance_evaluation}

\vspace{-0.1cm}
\subsection{Experimental Setup}

\textbf{Hardware:} All experiments are run on the IDRIS's Jean Zay supercomputer. Computation use an NVIDIA H100 80GB SXM GPU with driver 595. The host-driven batched configurations use one core of an Intel Xeon Platinum 8468 CPU.

\noindent\textbf{Software stack:} Our prototype is compiled with CUDA 13.0.3 and host GCC 14.2.0, in C++20. External solvers come from MAGMA 2.10.0, Kokkos-Kernels 5.1.0, cuSolverDx 0.4.0 with in-house patches, SNLS 0.4.4, and cuBLAS from CUDA 13.0.3 SDK.
The TFEL/MFront version is the master branch (May 2026). All computations are in double precision, with no fast-math optimizations.

\noindent\textbf{Compilation:} Device code is built with \texttt{--expt-relaxed-constexpr} and \texttt{-O3}. Solvers requiring cross-translation-unit device optimization additionally enable relocatable device code (\texttt{-rdc=true}) and device link-time optimization (\texttt{-dlto}). Host code is built with \texttt{-O3} and  \texttt{-DTFEL\_NO\_RUNTIME\_CHECK\_BOUNDS}.

\noindent\textbf{Kernel launch configuration:} Every in-house kernel selects its launch config through a uniform heuristic. To maximize occupancy and minimize tail effect, we retain the block size maximizing $O + W$, with $O$ the theoretical occupancy ratio and $W$ the average wave fill ratio. Batched solvers (cuBLAS, MAGMA) use their library-defined launch configuration unchanged.

\noindent\textbf{Measurement protocol:} Each measurement is repeated 11 times (one warm-up plus ten timed runs), and we report the median. We verify numerical correctness by comparison of the solution against a LAPACK reference solution. Detailed GPU profiling was performed using Nsight Compute.

\noindent\textbf{Reporting convention:} Some curves are labelled ``best X''. At each problem instance, this label reports the variant of solver family X with the smallest kernel duration. This reflects some realistic production cases, where a size-aware runtime dispatcher selects among specializations of the same kernel family.



\vspace{-10pt}
\subsection{Pure LUpp Performance}
\label{pure_lupp_performance}

Pure-LUpp performance is measured on batches of $10^5$ independent linear systems with two input distributions, neither of which produces singular matrices. The default draws matrix and RHS entries uniformly from $[-0.5, +0.5]$, mimicking the standard LINPACK benchmark configuration. A stress distribution instead draws from $[-5 \times 10^{-10},\ +5 \times 10^{-10}]$ to force small magnitudes and out-of-tile pivoting in all our tiled solvers, which all use a $10^{-10}$ pivot acceptance threshold. The default distribution produces no out-of-tile pivots. Under the stress one, depending on the tile size, roughly half the systems trigger at least one, averaging $1.5$ per triggering system. We quantify the stability cost (Section~\ref{tiledsolverdesign}) using the per-system normwise backward error (BE). Because pivoting is initially confined to a tile, BE grows as tiles shrink. A single tile recovers textbook LUpp and its BE order, while TileSize=2 gives the worst BE. Under the default distribution, the worst-case BE metrics across all matrix sizes are $10^{-15}$ (median), $10^{-14}$ (mean), and $10^{-9}$ (maximum). Under the stress distribution, out-of-tile pivoting widens pivot search and improves stability. Worst-case metrics then drop to $10^{-16}$ (median), $10^{-16}$ (mean), and $10^{-13}$ (maximum).

\begin{figure}[h!]
    \centering
    \includegraphics[width=1\linewidth]{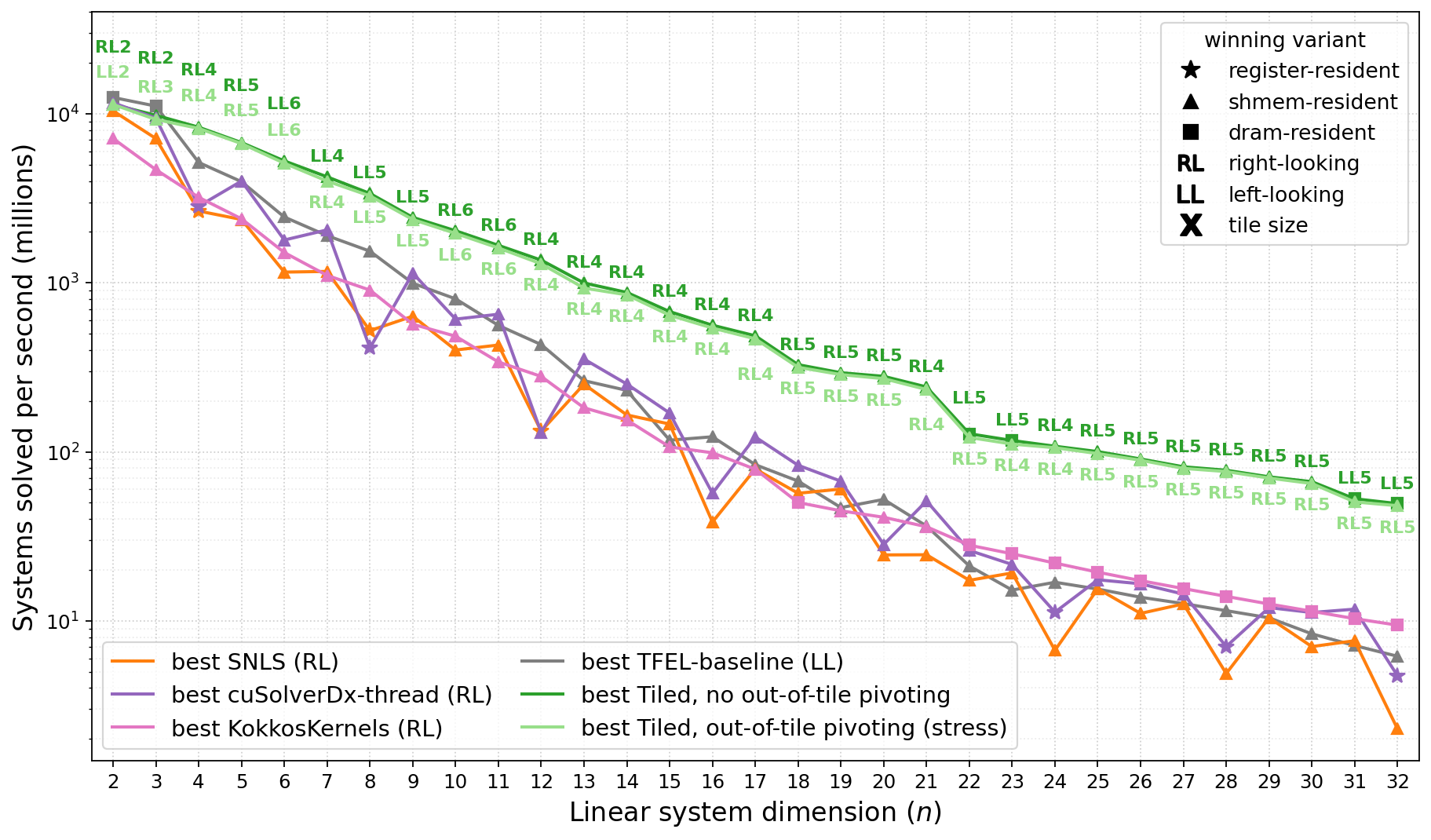}
    \caption{Throughput of the single-threaded LUpp solvers on $10^5$ linear systems (pure LUpp solving). Higher is better.}
    \vspace{-0.5cm}
    \label{fig:pure_lupp_singlethread_breakdown}
\end{figure}

\vspace{0.2cm}
Figure~\ref{fig:pure_lupp_singlethread_breakdown} shows single-threaded throughput across $n \in [2, 32]$. The in-house Tiled family dominates the $[4, 32]$ range, even under the stress distribution. All library-provided single-threaded solvers (SNLS, cuSolverDx-thread, KokkosKernels, TFEL-baseline) underperform: they conceptually stream the matrix scalar by scalar from their target residency into registers. Pivot indirections map directly onto memory accesses, and since pivot sequences differ across systems, these accesses are uncoalesced. Instead, tiled solvers improve data locality by caching full tiles in registers, and their static indexings guards prevent local memory usage~\cite{ref_local_memory_guard}. Without out-of-tile pivoting, their pivot indirections stay in registers, sparing them the uncoalesced accesses the library solvers incur. Across the $n \in [2, 32]$ range, the best Tiled-SHMEM variant issues a median $4\times$ fewer shared-memory loads and $18\times$ less local-memory traffic than TFEL-baseline in shared memory; the best Tiled-DRAM variant moves a median $3\times$ fewer DRAM bytes than TFEL-baseline in DRAM.

\vspace{0.2cm}
Under the default distribution, the residency of the winning tiled variant is selected by a trade-off between access cost and parallelism. Tiled-SHMEM keeps matrix accesses on-chip, but its shared-memory footprint lowers occupancy in steps: 4 warps/SM up to $n=14$, 3 from $n=15$, 2 from $n=18$, and 1 from $n=22$. Tiled-DRAM pays off-chip accesses but keeps 8 warps/SM throughout. Up to $n=21$, Tiled-SHMEM’s cheaper accesses dominate; from $n=22$, its limited parallelism erases this edge and both residencies perform within $\sim10\%$, with no consistent winner. The RL/LL choice follows the same trade-off and inverts with the residency. On Tiled-DRAM, LL wins because it reads each prior tile only when needed instead of repeatedly updating trailing tiles. On Tiled-SHMEM, shared-memory accesses are cheap, so LL gains little by reducing them. Meanwhile, it issues $\sim30\%$ more instructions than RL, which is why RL wins from $n \geq 14$. For $n \leq 13$, the matrix has too few tiles for either schedule to gain a measurable edge ($<5\%$). The optimal Tile Size (TS) follows the dimension while the matrix fits in one or two tiles $(n \leq 11)$, then locks to TS $\in \{4,5\}$, regardless of residency and distribution. Larger tiles amortize more updates per tile fetched, but register pressure caps the benefit (Section~\ref{tiledsolverdesign}).

Under the stress distribution, out-of-tile pivoting fires frequently and the winners shift. Tiled-SHMEM now dominates the whole range: on-chip access remains cheaper for the now-frequent out-of-tile pivot loads, and it keeps the lead in the sub-warp NTPB regime it enters from $n = 31$ (Section~\ref{subsubsec:single_threaded}), which stays shallow here ($\leq4$ idle lanes). A deeper sub-warp, at larger $n$ or with smaller shared-memory budgets, would favor Tiled-DRAM. The RL/LL choice shifts too: LL now wins only at some small sizes ($n \leq 10$); RL dominates from $n \geq 11$. This is because LL's out-of-tile pivot search must replay all prior $L \cdot U$ tile updates on each candidate row in the lower tiles, since LL has not materialized them into the trailing matrix. Since each candidate row replays every lower tile, this overhead grows with the tile count. RL avoids it because its trailing matrix is already updated, so candidate evaluation reduces to a raw load.

\begin{figure}[h!]
    \centering
    \includegraphics[width=1\linewidth]{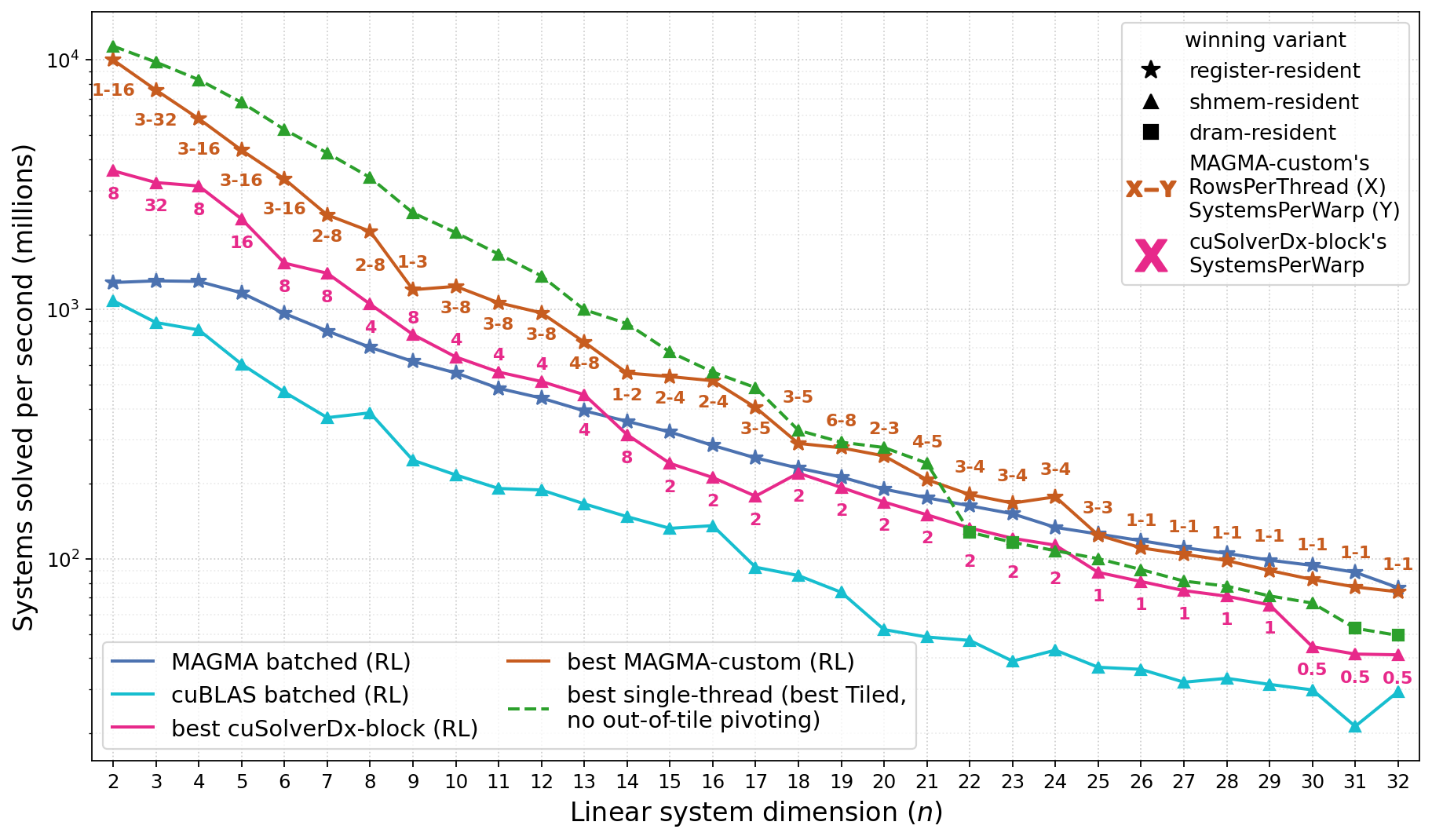}
    \caption{Throughput of the multi-threaded LUpp solvers on $10^5$ linear systems (pure LUpp solving). Higher is better.}
    \label{fig:pure_lupp_dim_sweep_oob}
    \vspace{-0.5cm}
\end{figure}

Figure~\ref{fig:pure_lupp_dim_sweep_oob} reports the multi-threaded throughput, with the best single-threaded Tiled curve from the previous figure overlaid as a cross-paradigm anchor. The most striking observation is that this single-threaded curve outperforms every multi-threaded solver for $n \in [2, 21]$, and remains competitive for the larger dims. It is overtaken by best MAGMA-custom and MAGMA only from $n = 22$. It always closely matches or beats cuSolverDx-block, and is always strictly faster than cuBLAS. In pure LUpp, over roughly the two thirds of the tiny regime, Tiled gains more from its data locality than multi-threaded solvers gain from their lower register pressure and higher occupancy.

Among multi-threaded solvers, best MAGMA-custom consistently outperforms best cuSolverDx-block, and outperforms MAGMA for $n \in [2, 24]$. The gain comes from MAGMA-custom's multiple rows per thread, and multiple systems per warp. Both eliminate the idle-lane waste of MAGMA's one-row-per-thread, one-system-per-block design. At $n = 25$, systems are too large for our packing optimizations to stay effective, and the two solvers track each other closely. From $n \geq 26$, the best MAGMA-custom variant uses RowsPerThread=SystemsPerWarp=1. It is therefore conceptually equivalent to the original MAGMA algorithm from this point on. The residual performance gap reflects minor implementation differences (generalization overhead and launch configuration), for an average $12\%$ throughput penalty.

Turning to the batched solvers cuBLAS and MAGMA: cuBLAS is consistently slower than MAGMA for $n \in [2, 32]$, confirming the benefit of register-blocking used by MAGMA. The latter is itself beaten by cuSolverDx-block for $n \in [2, 13]$, a range where MAGMA's idle-lane penalty is particularly high, with at least 19 of its 32 lanes idle. In this pure-LUpp setting, both batched solvers run as a single GPU kernel: they avoid the overheads they would incur in an application context such as the one studied in Section~\ref{subsec:full_constitutive_perf}.

In summary, thanks to its tile-granular register caching, the in-house Tiled family is the fastest option for $n \in [4, 21]$ under both default and stress distributions. From there, while Tiled remains competitive, MAGMA-custom takes the lead up to $n = 24$. Beyond, MAGMA narrowly takes over. The next section examines how these solvers perform in our applicative case.
\vspace{-0.4cm}
\subsection{Full Constitutive Law Evaluation Performance}
\label{subsec:full_constitutive_perf}
Full constitutive law evaluation performance is measured on a representative time step from a 3D simulation of the uniaxial compression of a nuclear fuel pellet, driven by an experimentally measured loading (compressive strain rate). The latter contains $10^5$ finite-element quadrature points, integrated under a Norton viscoplasticity model. Each of these integration points yields a Newton problem with $n=12$, with one linear system of this dimension to solve at every Newton iteration, plus one multi-RHS solve after convergence for the consistent tangent operator. Newton terminates on the first of the following: residual norm below $10^{-12}$ (always reached on this input), or 100 iterations (never reached). Convergence is highly homogeneous across the $10^5$ points: $2.1$ iterations on average, 3 at most. On this real data, Tiled solvers never block nor slow down Newton convergence. Their slight stability cost (Section~\ref{pure_lupp_performance}) thus carries no penalty in this applicative setting. Furthermore, they trigger no out-of-tile pivoting.

\begin{figure}[h!]
    \centering
    \includegraphics[width=1\linewidth]{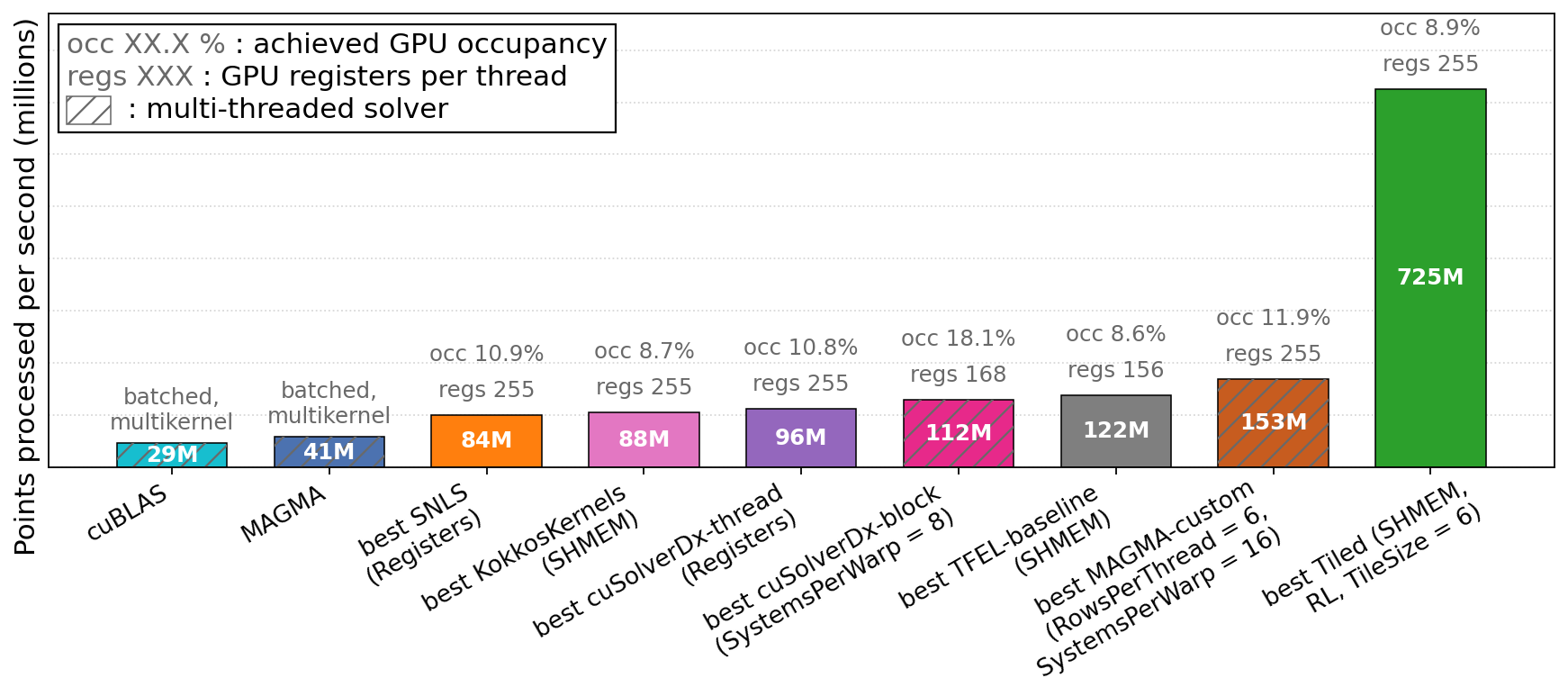}
    \caption{Throughput of all LUpp solvers on $10^5$ integration points (Norton constitutive law evaluation). Systems of dimension $n=12$. Higher is better.}
    \label{fig:repstep_classes}
    \vspace{-0.5cm}
\end{figure}

Figure~\ref{fig:repstep_classes} shows the throughput of every solver on this workload. The Tiled solver dominates by a wide margin. It is $4.7\times$ faster than the second-best solver (best MAGMA-custom). It reaches $5.9\times$ over the best library-provided device-callable competitor (TFEL-baseline), and $17.7\times$ over the best batched library (MAGMA). Strikingly, however, it exposes metrics typically associated with poor performance, with $8.9\%$ occupancy and $255$ registers per thread.
  
Multi-threaded device-callable solvers perform proportionally worse in this applicative setting than in pure LUpp: at $n=12$ in pure LUpp, the speedup of best Tiled over best MAGMA-custom was only $1.4\times$, and over best cuSolverDx-block was $2.6\times$. Here it grows to $4.7\times$ and $6.5\times$ respectively. Two factors compound. First, the multi-threaded paradigm pays additional synchronization barriers across the constitutive law evaluation. Second, the registers consumed by the Newton state and material-law variables raise local memory usage and push these solvers off the high-occupancy plateau they enjoyed in pure LUpp: best cuSolverDx-block drops from $48.2\%$ to $18.1\%$ achieved occupancy, and best MAGMA-custom drops from $23.8\%$ to $11.9\%$. The best Tiled solver is bottlenecked only by its shared-memory footprint, so its occupancy stays at $8.9\%$.

The batched solvers cuBLAS and MAGMA sit behind every device-callable alternative. In this applicative setting, they pay the paradigm overhead discussed in Section~\ref{existing_gpu_libraries_for_tiny_lupp}. One component of this overhead, the dragging of already-converged points, stays marginal here thanks to Newton's homogeneous convergence ($2.1$ iterations on average, $3$ at most). Both solvers would degrade further still on workloads with broader convergence dispersion.

\vspace{-0.4cm}
\section{Conclusion}
\label{sec:conclusion}
\vspace{-0.2cm}

Efficient LUpp solving is a key component for porting implicit constitutive law evaluation to GPUs. We benchmarked batched, multi-threaded device-callable, and single-threaded device-callable strategies for tiny LUpp, including two in-house solver families. Batched solvers are slowed down by applicative overheads and remain behind the TFEL library baseline. Among multi-threaded device-callable solvers, our MAGMA-custom solver improves on MAGMA library in pure LUpp for linear systems of dimension $n\in[2,24]$, but brings limited acceleration in the full constitutive workflow because of register pressure and synchronization costs induced by the multi-threaded approach. The strongest gains come from our single-threaded Tiled solver, which exploits tile-level register-caching. On the studied constitutive workload, this solver reaches $5.9\times$ over the TFEL baseline, $6.5\times$ over cuSolverDx, and $17.7\times$ over MAGMA.

Further work includes evaluating performance on newer GPUs with other backends and constitutive laws. Studying the Tiled solver data locality on CPU is another direction. Integrating it into TFEL/MFront (open-source) may follow, requiring only minor changes to MFront's code-generation logic.
\begin{credits}
\subsubsection{\ackname} This work was granted access to HPC computing and storage resources from GENCI at CNRS-IDRIS under grant 2026-101137, on the H100 partition of the Jean Zay supercomputer. \textbf{\discintname} The author has no competing interests to declare that are relevant to the content of this article.
\end{credits}

%
%
%
%

\end{document}